\documentclass[]{elsarticle}
\pdfoutput=1
\DeclareGraphicsExtensions{.pdf}

\biboptions{sort&compress}
\usepackage[colorlinks=true,linkcolor=blue, citecolor=blue, urlcolor=blue]{hyperref}
\usepackage{color}
\usepackage{multirow}
\usepackage{amssymb}
\usepackage{amsmath}
\usepackage{slashed}
\usepackage{graphicx}
\usepackage{placeins}
\usepackage{array}
\usepackage{dblfloatfix}
\usepackage{scalerel} 
\usepackage{subcaption}
\usepackage{multirow}

\begin{document}

\begin{frontmatter}
\title{
\center{Parametric analysis of 
COVID-19 expansion in European countries in the period of February to June 2020}
}

\author[Charles]{Martin Spousta \emailauthor{martin.spousta@cern.ch}{Martin Spousta}}
\address[Charles]{Institute of Particle and Nuclear Physics, Charles University, Prague, Czech Republic}

\begin{abstract}

The data on number of registered cases of COVID-19 disease in twenty European countries is analyzed by the least-squares fitting procedure with generic analytic 
functions. Three regimes of the expansion of the disease are identified and quantified -- early exponential expansion, damped exponential, and linear expansion. 
Differences among countries in the early expansion period are quantified. The velocity of the expansion in the exponential regime lies within one standard deviation 
from the average value for 11 countries. The number of infected individuals at the initial time is excessively high for Italy, 7 standard deviations from the average 
value.

Method for predicting the expansion based on extrapolation in the parametric space is presented. One week predictions based on extrapolations have average precision of 
18\% and 29\% during the later period of the damped exponential expansion for the case of Italy and Czechia, respectively. The method based on extrapolations in the 
parametric space may provide an elementary method to quantify the impact of restrictive measures on the spreading of the disease.

\end{abstract}

\begin{keyword}
Outbreak, Infection, SARS-CoV-2, COVID-19, Parametric analysis, European countries
\end{keyword}

\end{frontmatter}

\newlength{\fighalfwidth}
\setlength{\fighalfwidth}{0.49\textwidth}

\section{Introduction}
\label{sec:intro}

The outbreak of new coronavirus SARS-CoV-2, causing severe respiratory tract infection in humans, known as COVID-19, is a global health concern. Restrictive measures 
were adopted in many countries to mitigate the impact of the spread of the disease on the public health system \cite{who}.

The spread of an infectious disease is usually modeled using compartmental epidemiological models, see e.g. review in Ref.~\cite{Bauch2005} or examples of their recent 
implementation in Refs.~\cite{Bertuzzo2020,Hou2020,Prem2020}.
These models are often based on the numerical solution of a system of coupled linear differential equations for the number of susceptible (S), 
infectious (I), and recovered (R) cases (SIR models). 
For the simplest SIR model, a set of three differential equation characterizes the transitions among the three classes.

While it is not possible to obtain an analytic solution for a complex system of the SIR model equations, it is possible to construct generic analytic functions describing 
the data and quantify the robustness of such a description based on statistical tests. This approach is the backbone of this analysis. The main goal of this analysis is 
to identify and quantify robust features in the dynamics of spread of the disease and present a method that may help quantify the impact of restrictive measures on the 
spreading of the disease.

\section{Method and dataset}

In the simple SIR model, the equation which describes the change in the number of infectious cases reads $ \mathrm{d}I/\mathrm{d}t = (c_1 S/N - c_2) I$ 
where $N=S+I+R$ and $c_1, c_2$ are constants (see e.g. \cite{Hethcote2000} for details). 
  At the beginning of the outbreak 
when no restrictive measures are applied and the
susceptibility of the population to the virus is 100\%, one has 
$S \gg I$ and $R \rightarrow 0$. This leads to $dI/dt \propto I$ implying an exponential increase of the number of infected individuals with time,
\begin{equation}
I = I_0 \exp[ a t ], \label{eq:0}
\end{equation}
  where $I_0$ and $a$ are constants, combinations of $c_1, N$, and $S$. When restrictive measures are imposed, $c_1$ is not a constant but a function $c_1 = c_1(t, ...)$ 
introducing more complexity to the set of SIR model equations. Alternatively, the evolution under restrictive measures can be described by exponential dependence with a 
time-dependent parameter, that is, by the function
\begin{equation}
F(t) = N_0 \exp[ a(t) t ] = N_0 \exp[ b_1 t + b_2 t^2 + b_3 t^3 + ...], \label{eq:1}
\end{equation}
 where the $r.h.s.$ of (\ref{eq:1}) represents a Taylor expansion of the time dependent parameter $a$ and $N_0$ reflects number of infected individuals at $t=0$. Full 
description of the data can be achieved with $n$ $b_n$ parameters where $n$ is a large number. The $r.h.s.$ of (\ref{eq:1}) characterizes the basic exponential 
dependence plus any smooth deviation from this dependence.

  A large number of free parameters in (\ref{eq:1}) may be reduced and interpreted. At the beginning of the outbreak, 
  the number of cases is described by a pure exponential function ($b_1 \neq 0, b_{n>1}=0$) as argued in Eq.~(\ref{eq:0}). This hypothesis is tested in 
Section~\ref{sec:expo} and the parameters of the exponential function are compared among European countries. Once the non-zero $b_2$ is needed to describe the data, the 
expansion starts to slow down with respect to the purely exponential propagation. This is quantified in Section~\ref{sec:damping}. In an idealized case, only one $b_2$ 
coefficient along with previously fitted $b_1$ coefficient would be needed to describe the data. However, such a simple two-component characterization is shown to be 
invalid in Section~\ref{sec:damping}, and higher-order terms are needed or alternatively time-dependent $b_2$ and $N_0$ coefficients are needed to describe the data. 
The latter approach is used in this analysis. The evolution of parameters with time is studied, and predictability of the spread based on extrapolations is quantified. 
After a certain period, the description by a linear function outperforms the description by (\ref{eq:1}). The transition to a linear spread is quantified in 
Section~\ref{sec:linear}.

The cumulative number of registered cases of COVID-19 disease in European countries is used as input data to this analysis. Specifically, the analysis uses 
data from 20 European countries, 17 countries of the European Union 
exceeding 5 million of inhabitants and data from Norway, Switzerland, and the United Kingdom.
Data are taken from information system \cite{url1,Dong2020} for a period of 15$^\mathrm{th}$ of February to 1$^\mathrm{st}$ of June 2020. 
Analysis uses statistical tools implemented in {\it ROOT~6.1} \cite{Brun:1997pa}.

\section{Results}

\subsection{Period of exponential expansion}
\label{sec:expo}

During the first period, when there is no implementation of restrictive measures and the susceptibility of the population to the virus is 100\%, the number of all 
registered cases exponentially grows. This is described by Equation~(\ref{eq:1}) with $b_{n>1} = 0$ and $b_1$ quantifying the velocity of the spread of the disease, 
here in units day$^{-1}$. To quantify the expansion during this period we perform the least-squares fitting of the number of all registered cases by an exponential 
function in the time interval of $t_{min} - t_{max}$. The default lower bound $t_{min}$ was selected by a requirement of at least 50 cases to reduce fluctuations due to 
too small statistical sample. The upper bound $t_{max}$ was selected individually for each country based on comparing with an alternative description allowing for 
non-zero $b_2$ coefficient for each day when the number of cases exceeds 500. The comparison is based on a $\chi^2$ test. The value of $t_{max}$ represents a first day 
when fit with non-zero $b_2$ provides a better description than a fit with zero $b_2$. The resulting $b_1$ and $N_0$ coefficients are summarized in 
Table~\ref{tab:tab1}. The boundary condition leading to the definition of $t_{min}$ (50 cases) was varied by 20\% both up and down, which leads to alternative results. 
These alternative results were found to correspond with the nominal results within one standard deviation for 90\% of countries.

\begin{table} 
\begin{center}
\begin{tabular}{|c|c|c|c|} \hline
Country & $N_0$ & $b_1$ [day$^{-1}$] & $p_{KS}$ \\ \hline
Italy & 6.61 $\pm$ 1.02 & 0.39 $\pm$ 0.01 & 1.00 \\ \hline
Germany & 0.57 $\pm$ 0.13 & 0.35 $\pm$ 0.01 & 1.00 \\ \hline
Portugal & 0.01 $\pm$ 0.00 & 0.34 $\pm$ 0.01 & 1.00 \\ \hline
Sweden & 0.23 $\pm$ 0.06 & 0.31 $\pm$ 0.01 & 1.00 \\ \hline
Spain & 0.92 $\pm$ 0.20 & 0.30 $\pm$ 0.01 & 1.00 \\ \hline
France & 1.38 $\pm$ 0.39 & 0.30 $\pm$ 0.02 & 1.00 \\ \hline
Austria & 0.13 $\pm$ 0.03 & 0.30 $\pm$ 0.01 & 1.00 \\ \hline
Norway & 0.37 $\pm$ 0.14 & 0.29 $\pm$ 0.02 & 1.00 \\ \hline
Denmark & 0.27 $\pm$ 0.05 & 0.29 $\pm$ 0.01 & 0.94 \\ \hline
Czechia & 0.10 $\pm$ 0.03 & 0.27 $\pm$ 0.01 & 1.00 \\ \hline
Switzerland & 0.80 $\pm$ 0.11 & 0.27 $\pm$ 0.01 & 1.00 \\ \hline
Netherlands & 0.83 $\pm$ 0.15 & 0.26 $\pm$ 0.01 & 1.00 \\ \hline
Poland & 0.08 $\pm$ 0.03 & 0.25 $\pm$ 0.01 & 1.00 \\ \hline
Belgium & 0.94 $\pm$ 0.12 & 0.24 $\pm$ 0.00 & 1.00 \\ \hline
United Kingdom & 1.85 $\pm$ 0.37 & 0.22 $\pm$ 0.01 & 1.00 \\ \hline
Greece & 0.97 $\pm$ 0.19 & 0.19 $\pm$ 0.01 & 1.00 \\ \hline
Romania & 0.84 $\pm$ 0.16 & 0.18 $\pm$ 0.01 & 1.00 \\ \hline
Hungary & 0.26 $\pm$ 0.06 & 0.17 $\pm$ 0.01 & 1.00 \\ \hline
Bulgaria & 2.96 $\pm$ 0.37 & 0.11 $\pm$ 0.00 & 1.00 \\ \hline
Slovakia & 4.28 $\pm$ 0.07 & 0.10 $\pm$ 0.00 & 0.96 \\ \hline
\end{tabular} 
\end{center} 
\caption{ 
Parameters of fit ($N_0$, $b_1$) of the total number of cases in European countries during the first period of the expansion (for the definition see the text). Fit 
function is given by Equation~(\ref{eq:1}) with $b_i=0$ for $i>1$. The $p$-values for 
Kolmogorov-Smirnov test. Countries are ordered by a decreasing value of $b_1$ coefficient.
} 
\label{tab:tab1} 
\end{table} 

Table~\ref{tab:tab1} contains also results of Kolmogorov-Smirnov (KS) test. Based on $p$-values, one can 
accept the hypothesis that the exponential function provides a suitable description of the spread of the disease in the early stage for all the countries. 

The average value of $b_1$ is $0.25 \pm 0.04$ day$^{-1}$. The value of $b_1$ for 11 out of 20 countries lies within one standard deviation from the average value. 
The value of $N_0$ (which reflects the number of infected individuals at $t=0$) oscillates between zero and one for the majority of countries, but it is excessively high for the case of 
Italy (7 standard deviations from the average value calculated excluding Italy). This may be interpreted as evidence for the validity of a hypothesis that the spread in Italy has started sooner than when the first cases were reported and systematically traced (see e.g. Ref.~\cite{Tuite2020}).

To further check the universality of the initial velocity of the expansion quantified by $b_1$ coefficient,
we studied magnitude of Pearson's correlation coefficient quantifying the correlation between $b_1$ and other quantity $x$, $\rho(b_1, x)$. 
For $x$ we tested: 1) $N_0$; 2) number of days 
between 15th of February (first day when cases start to be collected European-wide \cite{url1}) and the day when the number of identified cases exceeds 50 in a given country ($d_{50}$); 3) 
the number of inhabitants in a given country $n_{inhab}$. The absolute value of the correlation coefficient was found to be smaller than 0.2 for all the quantities, implying that $b_1$ is not correlated with any of $N_0$, $d_{50}$, and $n_{inhab}$.

The value of $b_1$ can be further used as an input to specific epidemiological studies. In particular, it can be used to estimate the basic reproduction number, $R_0$, 
\cite{vandenDriessche2002} using Laplace transform as $R_0 = 1/[\int_0^\infty \exp(-b_1 \kappa) h(\kappa) \mathrm{d} \kappa]$ \cite{Zhao2019,Zhao2020}. Here $\kappa$ 
and $h(\kappa)$ represent the generation interval (i.e. the average time interval from onset of one individual to the onset of another individual) and its probability 
distribution, respectively.

\begin{figure*} [t]
\centering 
\includegraphics[width=0.89\textwidth]{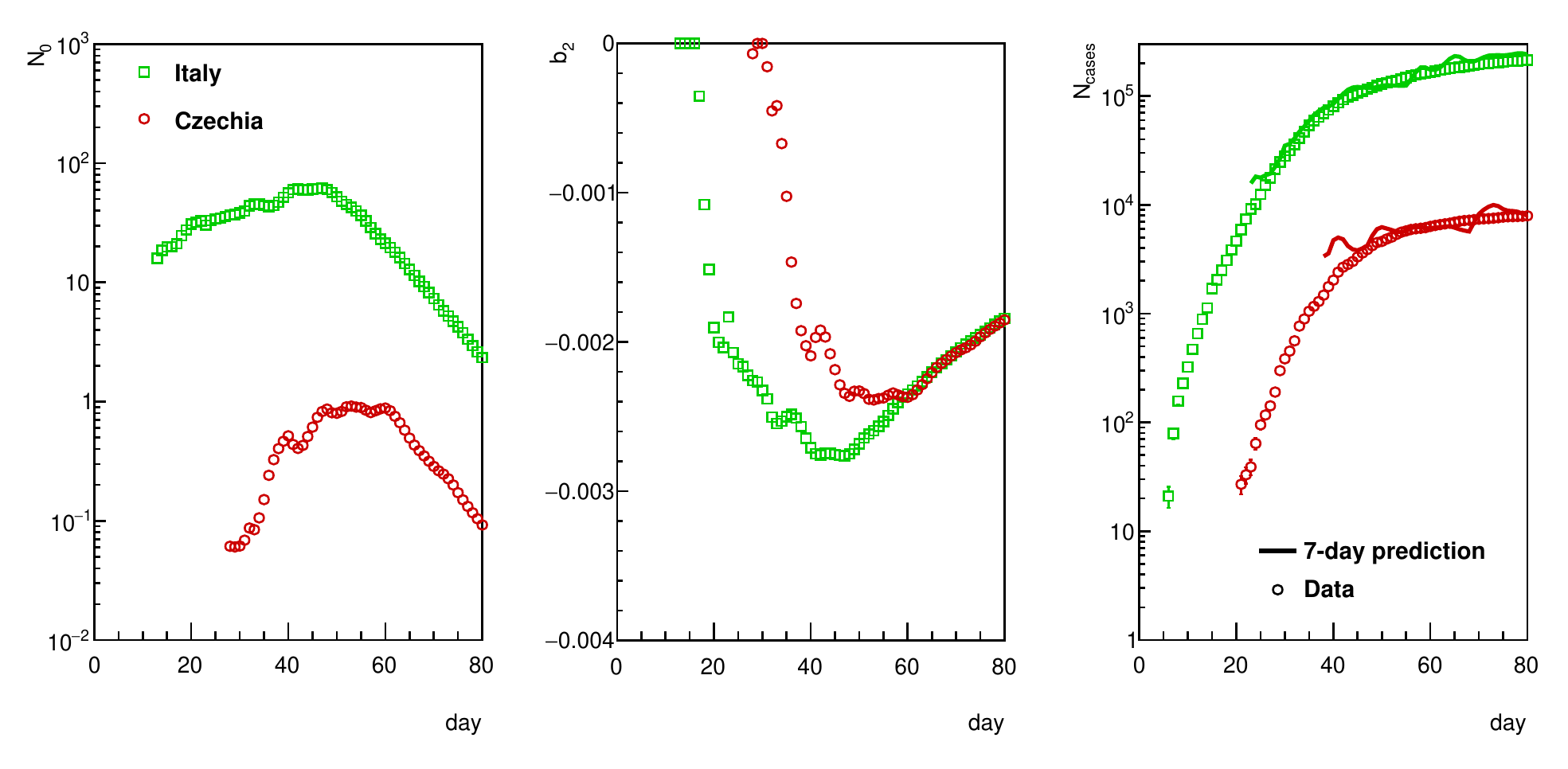} 
\caption{
The $N_0$ {\it (left)}, $b_2$ {\it (middle)}, and number of all cases {\it (right)} as a function of time for Italy {\it (green)} and Czechia {\it (red)}. The published data on the number of registered cases {\it (full markers)} are shown together with one-week predictions {\it (open markers)} (for details see the text).
} 
\label{fig:fig2}
\end{figure*}

\subsection{Period of exponential expansion with damping}
\label{sec:damping}

As shown in the previous section, the growth of the number of cases during the first period of spread of the disease can be characterized by an exponential function. 
The parameter $b_1$ can be interpreted as a parameter quantifying the uncontrolled spread of the disease. 
This first period ends when the fit starts to prefer a non-zero $b_2$. 
  Function from $r.h.s.$ of Equation~(\ref{eq:1}) with $b_i=0$ for $i>2$ is fitted to the number of all registered cases in a given country in a time window of 7 days. 
This time window moves in steps of one day to quantify the evolution of $N_0$ and $b_2$ with time.

While the complete description of the spread can be done by the function $a(t)$ from Equation~(\ref{eq:1}), here we take only the first two coefficients of the Taylor 
expansion of that function. We have tested that this provides a good description of the data ($p > 0.995$ from KS tests) in 7-day time window during the full second 
period. Allowing for non-zero $b_3$ or higher-order terms does not bring an improvement in the description. Therefore, one parameter, $b_2$, quantifies the degree of 
deviation from a purely exponential expansion in a given time. The more negative $b_2$ implies more significant ``damping'' of the original exponential growth. The 
units of $b_2$ used here are day$^{-2}$.

The example of time evolution of $N_0$ and $b_2$ for one large country (Italy) and one small country (Czechia) is shown in the left and middle panel of 
Figure~\ref{fig:fig2}. The parameter $b_2$ starts at zero (purely exponential expansion) and then rapidly decreases. The largest negative values are achieved for Italy. 
The $b_2$ distributions are qualitatively similar: first $b_2$ decreases fast, later the decrease is slower, ending at a minimum followed by a shallow increase where 
the $b_2$ achieves similar values for both countries.
While the interpretation of 
$b_2$ parameter is straigthforward, one can call it a ``push-back'' parameter, the interpration of time-dependent $N_0$ is not straightforward anymore and it is just a 
nuisance parameter.

In the case that $b_2$ turns to a constant in time, the evolution is predictable by simple analytic calculation. In particular the time $t'$ when no new cases should be 
registered, is given by a derivative of Equation~(\ref{eq:1}) with respect to time, leading to $t' = -b_1/(2b_2)$. Further, if $N_0$ is constant in time, number of all 
cases can be calculated by evaluating (\ref{eq:1}) at time $t'$. While $b_2$ and $N_0$ exhibit only a mild evolution during certain periods, these parameters do not 
converge to constant values. Prediction of the evolution of the value of registered cases cannot therefore be done analytically in general. 
  However, it can be done by extrapolating the evolution of parameters to the future. Extrapolations based on constant, linear, polynomial, and exponential functions 
were tested. Prediction from extrapolation was calculated for a value of registered cases in 7 days after a given time. These one-week predictions are shown along with 
the data in the right panel of Figure ~\ref{fig:fig2}. For the period when $b_2$ is decreasing, it was found that an extrapolation using constant values of parameters 
provides the best matching between estimates of the number of registered cases based on these extrapolations and the actual data. For the period when $b_2$ starts to 
increase, a linear extrapolation and exponential extrapolation of $b_2$ and $N_0$, respectively, provide predictions that are closest to the data.

For the case of Italy, the predictions differ from the data on average by $265\% \pm 160 \%$ during the first 14 days after the time when data start to prefer the 
parameterization with non-zero $b_2$. This substantial disagreement is given by a fast decrease of $b_2$ during the first two weeks which we did not succeed to describe 
universally by the extrapolation. For the next 50 days, the predictions differ from the data on average by $29\% \pm 19\%$ and $18\% \pm 10\%$ in the case of Italy and 
Czechia, respectively. 
These findings imply that the extrapolation of trends in the parameter space used in this study does not allow to make realistic predictions in the 
period of two weeks after the time when $b_2$ starts to be non-zero. However, for the later times, the method allows to make estimates with reasonable precision.

The method based on extrapolating the trends in the parametric space may provide an elementary method to quantify the impact of restrictive measures on the number of 
registered cases. This may be done by evaluating a difference between the extrapolation and the data for a time period when a given restrictive measure is applied. 

The fact that the method based on extrapolations provide a reasonable estimate of the future evolution for a period of one week is simple evidence for a significant 
inertia in the evolution.

\subsection{Period of linear evolution}
\label{sec:linear}

\begin{figure} [h]
\centering 
\includegraphics[width=0.44\textwidth]{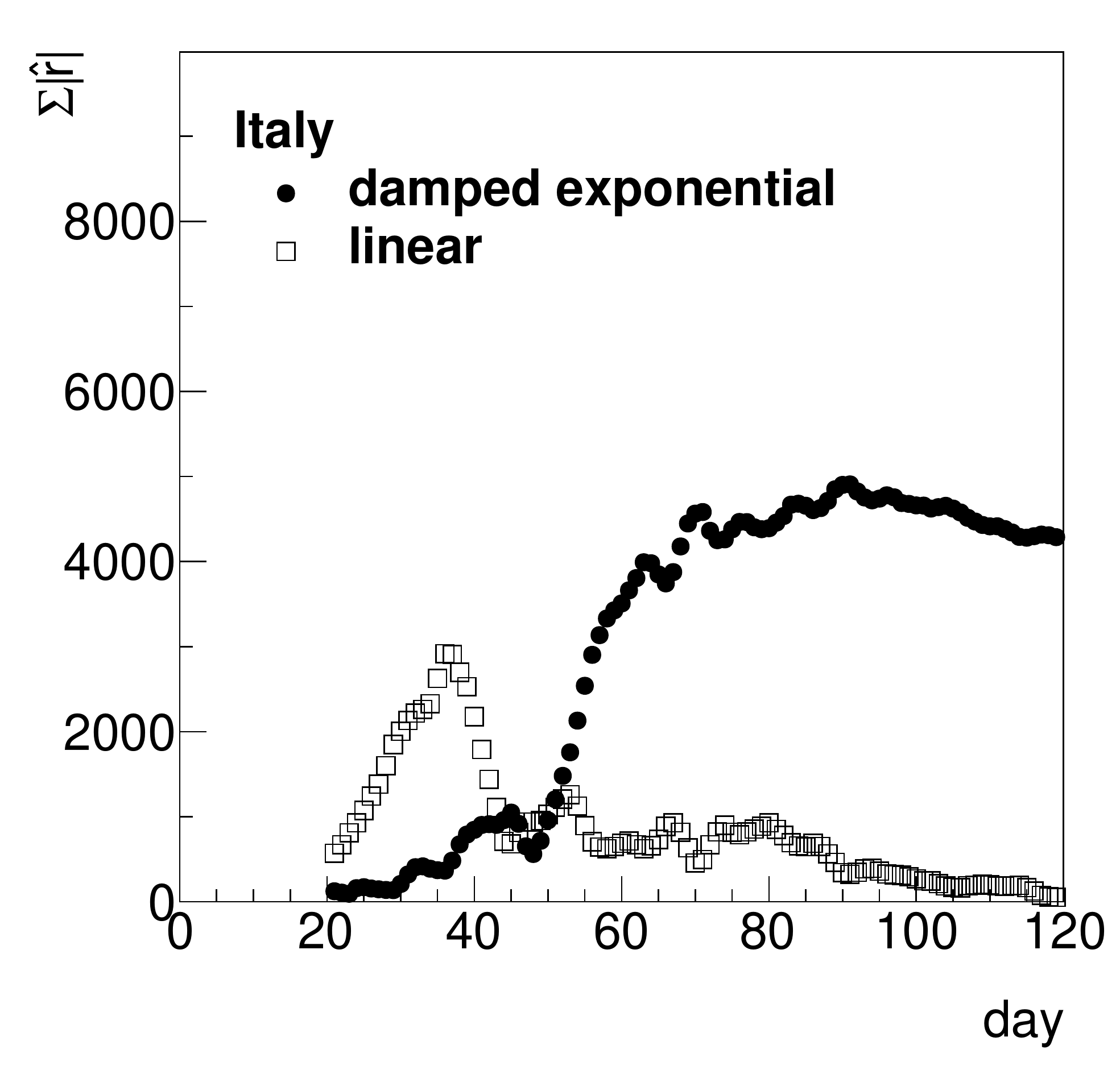} 
\caption{
Sum of fit residuals, $\sigma |\hat{r}|$, for the damped exponential and a linear fit of the number of all registered case evaluated as a function of time for Italy.
} 
\label{fig:fig3}
\end{figure}

The functional form of Equation~(\ref{eq:1}) stops describing the data well at some point. This reflects the fact that the spread of the disease turns from the damped 
exponential expansion to a linear evolution. This is quantified in Figure~\ref{fig:fig3} for the case of Italy in terms of the sum of residuals for the damped 
exponential fit and the linear fit. At around day 55, the linear function starts to characterize the data significantly better. The transition period between the damped 
exponential and linear growth takes about two weeks.

It is interesting to compare the value of $b_2$ for which the transition to linear evolution appears with the value of $b_1$. The transition occurs for $\sqrt{|b_2|}$ 
of $0.047 - 0.051$ day$^{-1}$ and $0.047 - 0.049$ day$^{-1}$ for Italy and Czechia, respectively. These values are roughly five times smaller than the average value of 
$b_1$ determined in Section~\ref{sec:expo}, $b_1 = 0.25 \pm 0.04$ day$^{-1}$.

It might be interesting to further explore if the transition to the linear evolution can be connected with some specific values of $b_2$ universally. If so, then the 
transition point may be predictable within the proposed parametric description of the spread. This is left open for further studies.

\section{Discussion}

Compartmental models represent an established way to model the spread of an infectious disease. As the COVID-19 pandemic evolves, they are being tuned and improved. 
In general, adding complexity into the description of the problem and adding more free parameters do not automatically guarantee a better precision of the description -- 
see e.g. discussion in Refs.~\cite{Jewell2020,Adam2020,Roda2020}.
  Work presented in this paper represents an alternative to complex compartmental modeling, which may provide new insight into the dynamics of the spread by its 
parametric analysis.

The spread of the disease is also often described by growth-law models, see e.g. Refs.~\cite{Zhao2019,Hsieh2009,Hsieh2017,Castorina2020}. Compared to the growth-law 
models, the parametric analysis presented here starts from generic analytic functions, a form of which is constrained by the data via statistical tests. Parameters are 
then let free to evolve. Trends in this evolution can be studied and extrapolated to future. This allows to make predictions and compare with the data which may provide 
an elementary method to judge the impact of restrictive measures.
  For example, if the data show a larger decrease compared to the extrapolation, then one gets an evidence that the restrictive measures applied earlier are efficient. 
This allows to attempt to compare the impact of different restrictive measures in a uniform way. This approach may be useful since the restrictive measures are 
complex and therefore hard to model. The undemanding approach used here may also allow to access larger datasets.

The cumulative number of registered COVID-19 cases from datasets used in this analysis represents a complex quantity that may be influenced by various external factors. 
In particular, these are the number of performed diagnostic tests and their sensitivity, number of asymptomatic cases, or simply a lack of accuracy (see e.g. discussion 
in Refs.~\cite{Bertuzzo2020,Bai2020,Adam2020}). This needs to be taken into account when interpreting results such as those presented in Table~\ref{tab:tab1}.

\section{Conclusions}
\label{sec:conclusions}

The analysis presented in this paper aims to add new information helping to understand the behavoir of the COVID-19 expansion in European countries. Data collected from 
20 European countries on number of registered COVID-19 cases was analyzed for a period of 15$^\mathrm{th}$ of February to 1$^\mathrm{st}$ of June 2020. Three regimes of 
the expansion are identified and quantified -- early exponential expansion, damped exponential, and linear expansion. Coefficient quantifying the velocity of the 
expansion in the early period ($b_1$) and coefficient quantifying the number of cases at time zero ($N_0$) were evaluated and compared for 20 European countries. The 
$b_1$ coefficient was found to lie within one standard deviation from the average value for 11 out of 20 countries. The $N_0$ for Italy was found to be excessively 
high, 7 standard deviations from the average value. The transition from the damped exponential regime to the linear regime is discussed and quantified in terms of the 
``push-back'' parameter ($b_2$).

A method to predict the evolution of the value of registered cases in the period of damped exponential expansion which is based on extrapolations
in parametric space was put forward. It was shown that it has an average precision of 18\% and 29\% in the period of $15-55$ days after the start of the damped
exponential expansion for the case of Italy and Czechia, respectively. This method may provide an elementary method to quantify the impact of restrictive measures on
the number of registered cases.

\section*{Acknowledgment}

The author would like to thank Ji\v r\' i Dolej\v s\' i, Dalibor Nosek, and Alfredo Iorio for useful discussions and careful reading of the manuscript.

\section*{Disclosure statement}

The author declares no conflict of interest. 

\section*{Data availability statement}

The data that support the findings of this study are openly available in Ref.~\cite{url1}.

\bibliography{document}
\bibliographystyle{elsarticle-num}
\end{document}